# Performance Analysis of Turbo Decoding Algorithms in Wireless OFDM Systems

Jorge Ortín, *Student Member*, IEEE, Paloma García, Fernando Gutiérrez and Antonio Valdovinos

**Abstract** — *Turbo codes are well known to be one of the error correction techniques which achieve closer results to the Shannon limit. Nevertheless, the specific performance of the code highly depends on the particular decoding algorithm used at the receiver. In this sense, the election of the decoding algorithm involves a trade off between the gain introduced by the code and the complexity of the decoding process. In this work we perform a thorough analysis of the different iterative decoding techniques and analyze their suitability for being implemented in the user terminals of new cellular and broadcast systems which are based on orthogonal frequency division multiplexing (OFDM). The analyzed iterative decoding algorithms are the Max-Log-MAP and the soft output Viterbi algorithm (SOVA), since both of them have a relative low computational complexity, simplifying their implementation in cost efficient terminals. Simulation results have been obtained for different encoder structures, block sizes and considering realistic channel conditions (an OFDM transmission over a wireless channel)[1].*

## I. INTRODUCTION

Turbo codes are one of the encoding techniques whose performance is closer to the channel capacity predicted by Shannon theorem. This fact has caused the adoption of turbo codes in diverse systems to perform forward error correction (FEC) strategies. For instance, Wimax, which is based on the 802.16e standard [1], and Long Term Evolution (LTE) [2], developed by the 3rd Generation Partnership Project (3GPP), have adopted turbo codes as the channel coding scheme for the transmission of user data in 4G mobile networks. Turbo codes are also employed in the DVB-RCS [3] standard for the return channel to satellite stations and the standards for broadcasting multimedia content to handheld terminals DVB-SH [4] and FLO [5]. Furthermore, recent research also proposes the use of turbo codes in Personal Area Networks (PAN) based on Mutiband OFDM for Ultra Wideband (UWB) [6]. Since all these technologies have very stringent power requirements, the use of turbo codes can be useful for improving the system performance and lowering the power consumption of the user terminals.

Turbo codes, as presented in [7], are formed by two parallel recursive systematic convolutional encoders separated by an inner interleaver at the transmitter and an iterative *A Posteriori Probability* (APP) decoder at the receiver. Several decoding algorithms to compute the APP of the encoded bits have been proposed up to now, which can be classified in two principal groups. One group is formed by those algorithms based on modifications of the Viterbi algorithm in order to obtain soft outputs which contain information related to the likelihoods of the decoded data. Amongst these algorithms, the suboptimal soft-output Viterbi algorithm (SOVA) is the most widely employed [8]. On the other hand, algorithms based on the optimal *Maximum a Posteriori* (MAP) algorithm try to decrease the computational load of this one at the cost of a penalty in the performance. Some of these algorithms are the log-MAP and the Max-log-MAP algorithm [9], whose specific implementation has also been optimized for real systems [10].

The election of the decoding algorithm will determine the effective correction capability of the code. Although the algorithms based on the MAP achieve best results than those based on the Viterbi algorithm, they also have a heavier computationally load. In this sense, it is necessary to evaluate the different available decoding strategies in order to find the algorithm which best fits to the requirements and limitations of user terminals. A correct election of the algorithm will maximize the performance of the code, decreasing the required SNR to achieve a specific error rate and thus reducing the cost of the receiver front end and the power consumption of the transmitter, while limiting the complexity of the algorithm and hence decreasing the cost of the associate digital electronic to implement it. These two parameters are of paramount importance in order to obtain cost effective terminals which will allow the broad spread of these new services.

The objective of the proposed work is to evaluate and compare the different iterative decoding techniques which can be used in the systems where it has been proposed the use of turbo codes considering the carrier technology and the transmission conditions found in these systems. In this sense, the different decoding schemes have been simulated considering the transmission of an OFDM signal whose parameters are similar to those employed in these systems over a realistic mobile radio channel with frequency-selective fading and Doppler spread. Similar analyses have been performed recently for other coding strategies implemented in IPTV systems [11].

The rest of the paper is organized as follows. Section II explains the structure of the turbo encoders employed in the

[1] This work has been financed by the Spanish Government (Project TEC 2008-06684-C03-02/TEC), Gobierno de Aragón (Project PI003/08 and WALQA Technology Park) and the European IST Project EUWB.

The authors are with the Aragón Institute for Engineering Research (I3A), University of Zaragoza, Zaragoza, 50018, Spain (e-mails: jortin@unizar.es, paloma@unizar.es, ferguso@unizar.es, toni@unizar.es).

evaluation of the algorithms. Section III describes the constituent blocks of the decoder and the iterative decoding process. Section IV presents and compares the simulation results obtained with the analysed encoders and decoders. Finally, section V concludes the paper.

## II. TURBO ENCODERS STRUCTURE

The inner structure of the turbo encoder is formed by two parallel convolutional encoders separated by an interleaver. The output of this block is usually followed by a puncturing block to adapt the code rate to the required one. The component encoder, the interleaver and the puncturing block are the principal parts of the encoder which influence the performance of the codes. In Figs. 1 and 2 the structure of two different convolutional turbo encoders without the puncturing block are illustrated.

The component encoder of a turbo code is a recursive systematic convolutional encoder, which can be binary or non-binary. Since the performance of any convolutional code relies on the free distance and its multiplicity, the component encoders should have maximum effective free distance and minimum multiplicity to achieve good results.

For instance, the component encoder presented in Fig. 1 is a 1/2 binary encoder with constraint length 4 and generator polynomials $1+D+D^3$ for the parity bit and $1+D^2+D^3$ for the feedback branch. In contrast, the component encoder of Fig. 2 is a double binary 1/2 encoder with constraint length 4 and generator polynomials $1+D^2+D^3$ and $1+D^3$ for the parity bits and $1+D+D^3$ for the feedback branch. In this kind of encoders two consecutive bits are fed into the encoder simultaneously, decreasing the latency of the encoding of each block. Besides, this kind of encoders have better convergence, larger minimum distances and less sensitivity to puncturing patterns than binary codes [12].

In order to preserve the distance properties of the code at the end of each block, a proper termination of the trellis is required. One possible solution is to feedback the final state of the encoder to its input to reset the encoder memory. This solution requires the transmission of these tail encoded bits to the receiver, which decreases the code rate. Another approach is the use of tailbiting. In this case, the encoder memory is initialized with the same state that it will end after encoding, avoiding the overhead of transmitting the tail.

The interleaver has influence in the code properties and in the iterative decoding performance. The interleaver breaks the input sequence prior to the second encoding thus increasing the code free distance when the output of both encoders are jointly considered. In this sense, the interleaver allows constructing codes with good distance properties from short memory convolutional codes. The interleaver also spreads burst errors to the second component decoder so that an iterative decoding based on uncorrelated information exchange between the two component decoders can be applied.

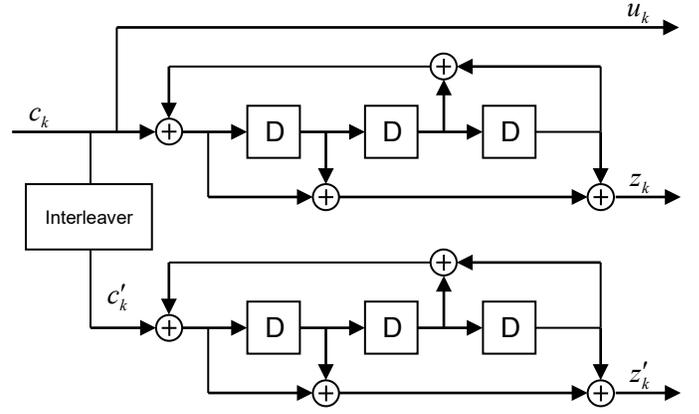

Fig. 1. Binary convolutional turbo encoder with generator polynomials $1+D+D^3$ for the parity bit and $1+D^2+D^3$ for the feedback branch.

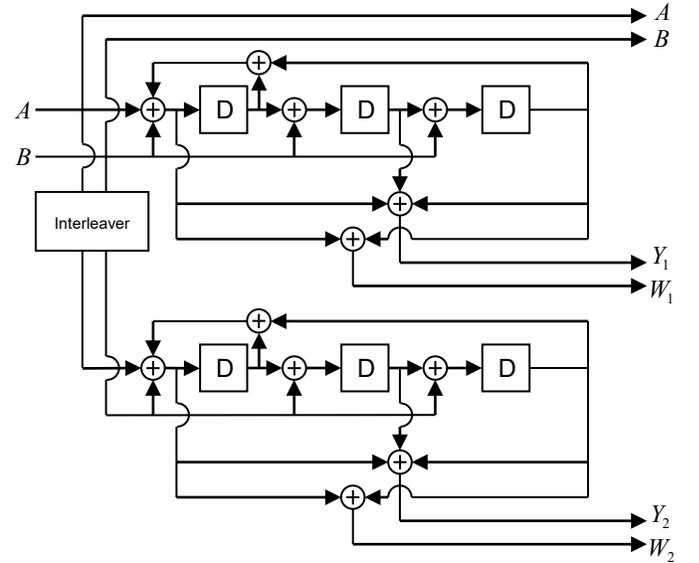

Fig. 2. Double binary convolutional turbo encoder with generator polynomials $1+D^2+D^3$ and $1+D^3$ for the parity bits and $1+D+D^3$ for the feedback branch.

Typically, the performance of a turbo code is improved when the interleaver size is increased, which explain the better performance of turbo codes with longer encoding blocks.

The output of the constituent encoders generally has a low rate and it is necessary to adapt it depending on the channel conditions. For this purpose, the encoder is followed by a puncturing block which alternately selects the outputs of the constituent encoders to form an encoded block with the required rate.

Since the aim of this work is to evaluate the different decoding algorithms which could be implemented in the user terminals of the new OFDM systems, the simulations will gather results obtained with two different turbo encoders to span different possible scenarios deployed in real wireless systems. The chosen encoders are the proposed one for its use in Wimax, whose constituent encoder is a double binary tailbiting code (the same constituent encoder is employed in the DVB-RCS standard) and the turbo code of the LTE standard, with a binary code with the final state feed backed to terminate the trellis as its constituent encoder. A similar code is also employed in the DVH-SH and the FLO standards.

## III. DECODING ALGORITHMS

An iterative turbo decoder is formed by two concatenated component decoders separated by the interleaver used in the encoder. The decoding algorithms must be Soft Input/Soft Output (SISO) algorithms, since they must accept *a priori* information at their input and produce *a posteriori* information at their output. Similarly to convolutional codes, trellis based algorithms such as the MAP and the Viterbi algorithm are used to compute this *a posteriori* information.

The principle of operation of an iterative decoder is based on the repeated exchange of estimates of the information (systematic) bits between the two SISO decoders as depicted in Fig. 3. The SISO algorithm provides a soft estimation of the coded bits measured in terms of their log-likelihood ratios, which for a binary random variable $u_k$ is defined as:

$$L(u_k) = \ln \frac{P(u_k = +1)}{P(u_k = -1)} \quad (1)$$

where $u_k$ is the information bit at time $k$. Each SISO decoder takes the soft estimations provided by the demodulator, which corresponds to the received systematic ($L_C \cdot y_I$) and parity bits ($L_C \cdot y_P$) together with the *a priori* estimations ($L(u)$) of the information bits provided by the other SISO decoder in the previous iteration.

The output of the SISO algorithm is formed by the *a posteriori* log-likelihood ratio of all information bits ($L(\hat{u})$) and their corresponding extrinsic information $L_e(\hat{u})$, which corresponds for each information bit to the soft output information provided by all other coded bits without the influence of the $L(u)$ and $L_C \cdot y_I$ values of that bit:

$$L(\hat{u}) = L_C \cdot y_I + L(u) + L_e(\hat{u}) \quad (2)$$

Thus, the estimation of each information bit is composed of three terms: the received value of this bit provided by the demodulator, the *a priori* information provided by the other decoder and the information obtained from the rest of the received coded sequence. The last term corresponds to the extrinsic information and is used by the other decoder as *a priori* information.

In the first iteration, the first decoder calculates the extrinsic information isolating the term $L_e(\hat{u})$ in (2) and substituting the *a priori* information with the extrinsic information provided by the other decoder:

$$L_e^1(\hat{u}) = L^1(\hat{u}) - L_C \cdot y_I - L_e^2(\hat{u}) \quad (3)$$

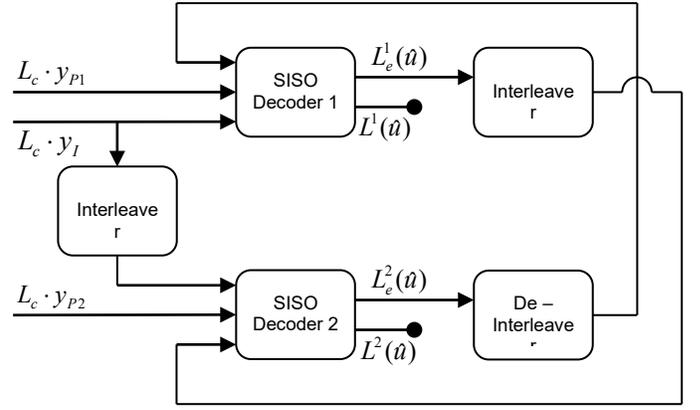

Fig. 3. Structure of an iterative turbo decoder.

Since in the first iteration there are not values provided by the decoder 2, this value is set to zero. Then, the extrinsic information computed by the first decoder is passed to the second decoder, which uses it as *a priori* information. The extrinsic information computed by the second decoder is:

$$L_e^2(\hat{u}) = L^2(\hat{u}) - L_C \cdot y_I - L_e^1(\hat{u}) \quad (4)$$

This extrinsic information is passed to the first decoder, which performs again a new decoding. This computation is repeated iteratively until a terminating condition is fulfilled. The final decoded sequence is based on the sign of the *a posteriori* log-likelihood ratios provided by the second decoder in the last iteration.

The optimal SISO algorithm to compute the likelihoods of the data bits $u_k$ is the MAP algorithm. This algorithm estimates the probability of the different information bits being +1 or -1 given the received coded sequence, and hence can easily compute the log-likelihood ratios used in the iterative turbo decoding. Nevertheless, the use of probabilities whose numerical representation leads to resolution problems and the large amount of exponentials and multiplications required to compute these probabilities makes this algorithm difficult to implement in real systems.

As a result, different approximations of this algorithm such as the log-MAP and the Max-log-MAP algorithms have been proposed to decrease the computational complexity of the MAP algorithm. These algorithms work in the logarithmic domain and hence transform the multiplications into additions, which are easier to compute. Out of these algorithms, the Max-log-MAP is the simpler in terms of computational complexity, although it presents some degradation in the performance.

Other algorithms which can obtain soft estimations of the data bits are those based on the Viterbi algorithm, such as the SOVA. This algorithm has a reduced complexity, slightly higher than that of the Viterbi algorithm and lower than those of the MAP-based algorithms, including the Max-log-MAP algorithm. Nevertheless, the expected performance of this algorithm is also lower.

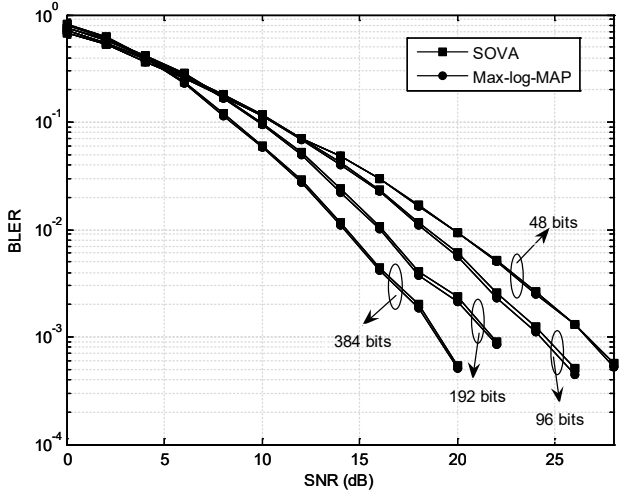

Fig. 4. BLER versus SNR for the binary turbo code in the Ped A channel with QPSK modulation, 1/2 code rate and different block sizes.

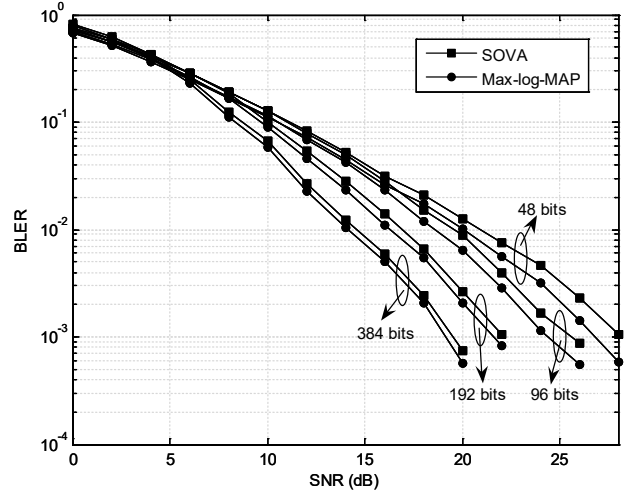

Fig. 5. BLER versus SNR for the duo-binary turbo code in the Ped A channel with QPSK modulation, 1/2 code rate and different block sizes.

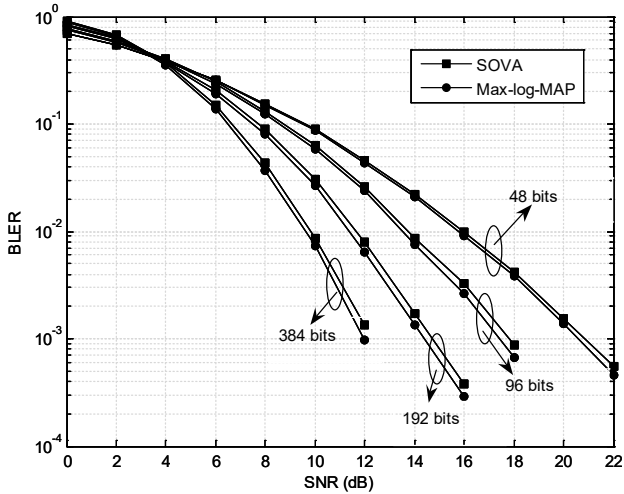

Fig. 6. BLER versus SNR for the binary turbo code in the Veh A channel with QPSK modulation, 1/2 code rate and different block sizes.

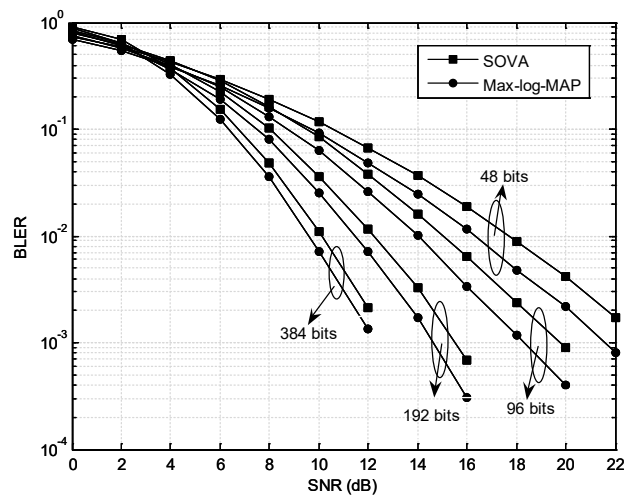

Fig. 7. BLER versus SNR for the duo-binary turbo code in the Veh A channel with QPSK modulation, 1/2 code rate and different block sizes.

Since the objective of the present work is to evaluate the different alternatives to decode turbo codes in user terminals of wireless OFDM systems, the considered algorithms to perform the analysis are only those with the lowest computational complexity, i.e., the Max-log-MAP and the SOVA.

## IV. PERFORMANCE EVALUATION

In this section, we show the performance of the Max-log-MAP and the SOVA algorithms in an OFDM system over the extended ITU Pedestrian A and ITU Vehicular A channels [13]. These models correspond to a fading multipath channel in an urban environment adapted to an OFDM transmission at pedestrian and vehicular speeds, which in these simulations are set to 3km/h and 30km/h, respectively. This choice aims at simulating the performance of the decoding algorithms considering the transmitting technology and channel conditions which will affect the user terminals of OFDM systems in normal conditions.

The considered codes are the binary code presented in Fig. 1 and the double binary code depicted in Fig. 2. Although these constituent encoders are shared amongst several standards as noted in section II, the specific interleavers and puncturing patterns are the ones defined in the LTE and the Wimax standards respectively. The sizes of the encoded blocks are variable.

The specific implementations of the Max-log-MAP and the SOVA algorithms for the binary code are detailed in [9] and [8]. The implementation of the Max-log-MAP for the double binary code is straightforward. The modifications required to use the SOVA algorithm in non-binary codes are detailed in [13]. The number of iterations employed in the decoding process is set to 8.

The simulated OFDM signal has a bandwidth of 5 MHz, the carrier frequency is 2.5 GHz and the number of subcarriers per symbol is 512. These parameters are compliant with both the

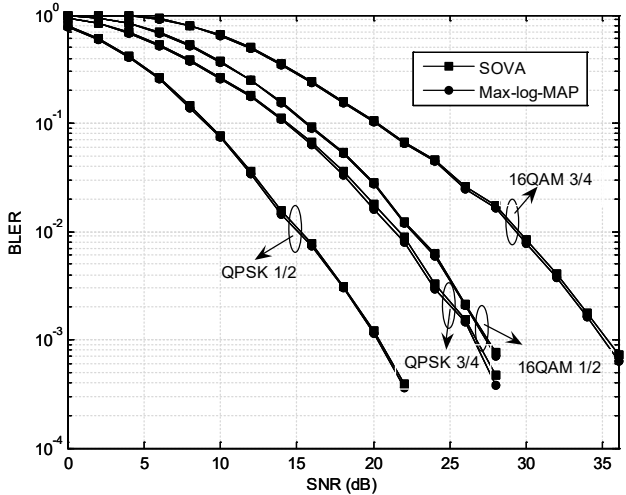

Fig. 8. BLER versus SNR for the binary turbo code in the Ped A channel with a block size of 288 bits and different modulations and code rates.

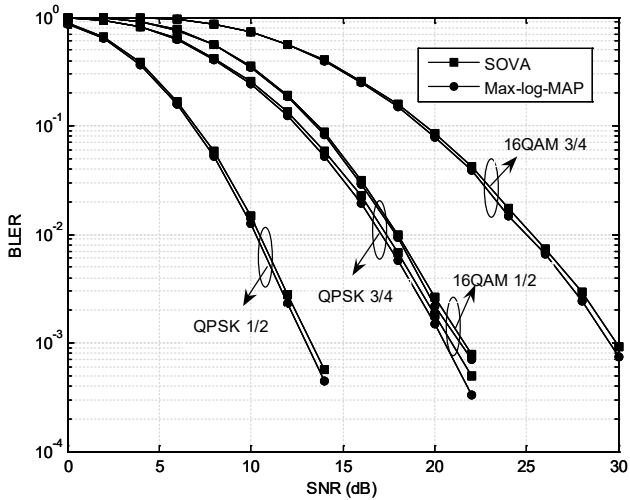

Fig.10. BLER versus SNR for the binary turbo code in the Veh A channel with a block size of 288 bits and different modulations and code rates.

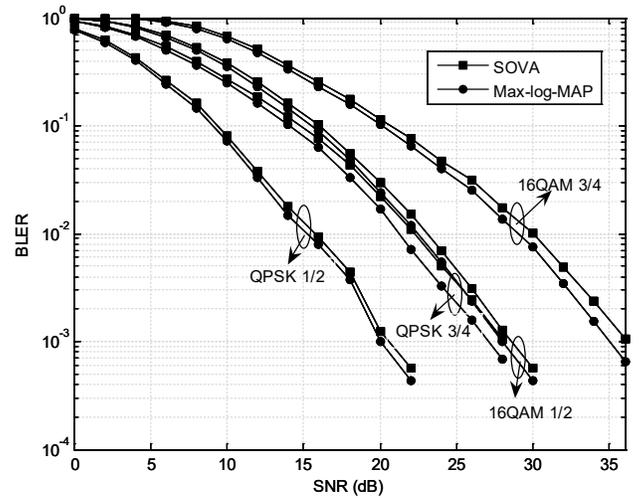

Fig. 9. BLER versus SNR for the duo-binary turbo code in the Ped A channel with a block size of 288 bits and different modulations and code rates.

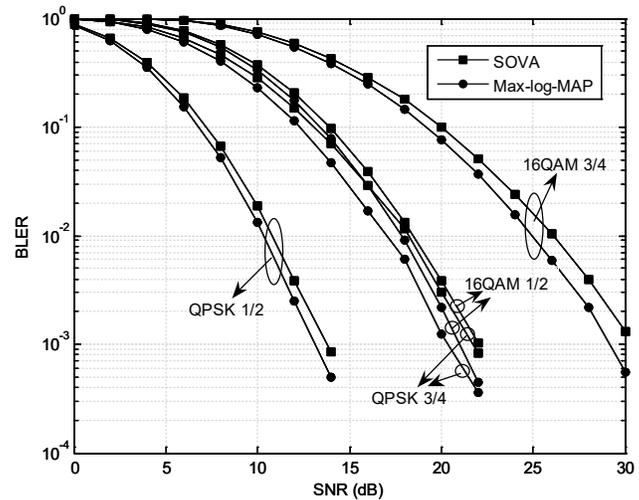

Fig.11. BLER versus SNR for the duo-binary turbo code in the Veh A channel with a block size of 288 bits and different modulations and code rates.

Wimax and the LTE standards and similar to those which can be found in other OFDM systems. The encoded data are interleaved and mapped to QPSK or 16QAM symbols. It has been assumed ideal channel estimation at the receiver, where Zero Forcing equalization is performed. The results are presented in terms of Block Error Rate (BLER) instead of BER since BLER represents a more accurate measure of the system performance as the ARQ retransmissions and the actual data throughput are function of this measure. The results have been obtained using Monte Carlo method, ensuring that at least 100 errors have been reported for each simulation result.

Fig. 4 and Fig. 5 show the performance of the binary and non-binary turbo codes over the ITU Pedestrian A channel for QPSK modulation, 1/2 code rate and different block sizes. The improvement in the performance of the Max-log-MAP over the SOVA is negligible for the binary turbo code and ranging from 0.2 to 1.5dB for the double binary turbo code. This penalty in the BLER for the non-binary code is caused by the tailbiting initialization of the constituent encoder, which worsens the performance of the Viterbi decoding algorithms for short block sizes.

Similar results are depicted in Fig. 6 and Fig. 7 for the ITU Vehicular A channel. In this case, the penalty for the use of the SOVA algorithm is 0.2dB respect to the Max-log-MAP for the binary turbo code and nearly 2dB for the non-binary code. The improvement in the performance obtained in this channel is due to the higher diversity that presents the carriers of the OFDM signal.

In Fig. 8 and Fig. 9 the performance of the binary and non-binary turbo codes over the ITU Pedestrian A channel for different modulation and code rates with a fixed block size of 288 bits is depicted. Again, the improvement in the performance of the Max-log-MAP over the SOVA is negligible for the binary turbo code and small for the double binary turbo code. These results are similar when the simulations are carried out over the ITU Vehicular A channel, as it is shown in Fig. 10 and Fig. 11.

The computational complexity of both algorithms depends on the number of iterations required to obtain a stable output and the additions and comparisons performed per iteration. According to the obtained simulation results, the number of iterations until a stable output is achieved is similar for both decoding algorithms and independent of the turbo encoder employed and the channel conditions.

The additions and comparisons required by each decoding algorithm are shown in Table 1. These listed operations correspond to the additions and comparisons performed in each pass of the algorithm for decoding a block of 384 data bits. It must be noted that these results do not enclose the operations required for the pass of information between decoders (the calculations of the extrinsic information from the *a posteriori* log-likelihood ratios) and that two passes of the decoding algorithm are performed per iteration. It can be seen that for the binary code, the computational complexity of the SOVA is much lower than that of the Max-log-MAP, whereas for the double binary code they are similar.

To summarize, the obtained figures in terms of BLER and computational complexity show that the SOVA algorithm is preferred over the Max-log-MAP when a binary turbo code is used, while the Max-log-MAP achieves better results with the double binary turbo code. Finally, it can be noted that there is not appreciable difference in the performance of both encoders, while the complexity associated with the decoding of the binary code is notably lower.

TABLE I
COMPARISON OF THE REQUIRED NUMBER OF OPERATIONS IN THE DECODING ALGORITHMS (DATA BLOCK SIZE: 384 BITS)

| Algorithm | Additions | Comparisons |
|---|---|---|
| SOVA (binary code) | 7353 | 4053 |
| Max-log-MAP (binary code) | 20124 | 12384 |
| SOVA (double binary code) | 17856 | 14897 |
| Max-log-MAP (double binary code) | 24000 | 15360 |

## V. CONCLUSION

The performance of different iterative decoding algorithms for turbo codes in new wireless systems is evaluated in realistic radio conditions, i.e., considering the transmission of an OFDM signal over a fading channel. The evaluation has been carried out with two different encoders, being one of them non-binary, in order to obtain representative results valid for other possible encoders which can be used in OFDM systems. The simulations have been conducted for various block sizes, code rates and modulations and the comparison between the algorithms is presented not only in terms of error rates, but also considering their computational complexity. Simulation results for the binary turbo code show that the SOVA achieves a performance similar to that of the Max-log-MAP algorithm but with a considerable lower computational load.

For the double binary encoder, the closer computational complexity of both algorithms and the gain of 2dB achieved with the Max-log-MAP make this algorithm preferred over the SOVA.

BIOGRAPHIES

**Jorge Ortín** was born in Zaragoza, Spain, in 1981. He received the Engineer of Telecommunications degree from the University of Zaragoza in 2005 and he is currently a Ph.D. student at University of Zaragoza, working in the area of mobile systems. In 2008 he joined Institute of Engineering Research in Aragon (I3A) of University of Zaragoza, where he has participated in different projects funded by public administrations and by major industrial and mobile companies. Research interests include wireless communications systems, with emphasis on channel coding and HARQ strategies.

**Paloma García-Dúcar** was born in Zaragoza, Spain, in 1972. She received the Engineer of Telecommunications and Ph.D. degrees from the University of Zaragoza (Spain) in 1996 and 2005, respectively. In 1995 she was employed in TELTRONIC SAU where she worked in the Research and Development Department, involved in the design of radio communication systems (mobile equipment and base station) until 2002. Since 1997 until 2001, she has collaborated in several projects with the Communication Technologies Group of the Electronics Engineering and Communications Department in the University of Zaragoza. In 2002, she joined the Centro Politécnico Superior, University of Zaragoza, where she is an Assistant Professor. She is also involved as researcher with the Aragon Institute of Engineering Research (I3A). Her research interests are in the area of linearization techniques of power amplifiers, and signal processing techniques for radio communication systems.

**Fernando Gutiérrez** was born in Barcelona (Spain) on 1966. He received the Engineer of Telecommunications MS from the Universitat Politècnica de Catalunya (UPC), Spain, and Ph.D. degrees from the Universidad de Zaragoza, Spain, in 2000. In 1993, he joined the Universidad de Zaragoza as Assistant Professor. He is member of Institute of Engineering Research in Aragon (I3A). His professional research activity lies in the field of wireless communications.

**Antonio Valdovinos** was born in Barbastro, Spain, in 1966. He received the Engineer of Telecommunications and Ph.D. degrees from the Universitat Politècnica de Catalunya (UPC), Spain, in 1990 and 1994, respectively.

In 1991 he joined, under a research grant, the Signal Theory and Communications Department, (UPC), where he was an Assistant Professor until 1995. In 1995 he joined the Centro Politécnico Superior, Universidad de Zaragoza (Spain) where he became an Associate Professor in 1996 and a Full Professor in 2003. He is a member of the Aragon Institute of Engineering Research (I3A). At present his research interest lies in the area of wireless communications with special emphasis on Packet Radio Networks, Wireless Access Protocols, Radio Resources Management and QoS.